\documentclass[11pt]{article}

\usepackage[margin=1in]{geometry}
\usepackage[T1]{fontenc}
\usepackage{lmodern}
\usepackage{microtype}
\usepackage{amsmath}
\usepackage{amssymb}
\usepackage{amsthm}
\usepackage{booktabs}
\usepackage{graphicx}

\newtheorem{definition}{Definition}
\newtheorem{proposition}{Proposition}
\usepackage{enumitem}
\usepackage{xcolor}
\usepackage[hidelinks,colorlinks=true,linkcolor=blue!50!black,
            urlcolor=blue!50!black,citecolor=blue!50!black]{hyperref}
\usepackage[capitalize,noabbrev]{cleveref}
\usepackage[numbers,sort&compress]{natbib}
\usepackage{array}
\usepackage{placeins}
\usepackage{capt-of}
\usepackage{pgfplots}
\pgfplotsset{compat=1.18}

\title{\bfseries An Application-Layer Multi-Modal Covert-Channel
       Reference Monitor for LLM Agent Egress}

\author{%
  Alfredo Metere\\
  Enclawed, LLC, California, USA\\
  \texttt{alfredo.metere@enclawed.com}
}
\date{\today}

\begin{document}

\maketitle

\begin{abstract}\noindent
A large language model (LLM) agent that sends messages can leak
data \emph{inside} them. Destination allowlists and content
scanners do not police whether an otherwise-benign payload is
itself a covert channel: a compromised agent encodes bits in
zero-width characters, homoglyphs, whitespace, base64, JavaScript
Object Notation (JSON) key ordering, message timing or size ---
and, in binary egress, in least-significant-bit (LSB) pixel
planes, per-image mean luminance, inter-image sequence
permutation, ultrasonic tones, or audible-band sonified data. Our
egress reference monitor has three contributions. (i)~A text
pipeline of ten capacity-reducing stages, a per-sink leaky-bucket
capacity ledger, and a staged posture that enforces lossless
stages from day one. (ii)~Two media scramblers (a Fourier-domain
audio band-limiter and a red-green-blue (RGB) image bit-depth and
mean-luminance bucketer) gated by a \emph{boot-time cryptographic
legitimacy attestation}: an auditor publishes at boot the trusted
Ed25519 keys and $\{$kind, data-class$\}$ pairs; only payloads
with a verifying signature for an authorized class are exempt.
The attestation sidesteps the intractable content-based
discrimination between real media and data sonified or rasterized
as a carrier; unsigned media is suspect by default; a
content-addressed canonicalizer closes the inter-image
permutation channel. (iii)~Residual capacity is the Miller--Madow
corrected mutual information between embedded and recovered bits
(zero when destroyed), measured by an adversarial ensemble of
fifteen working encoders across text, image and audio. The
reference implementation drives residual capacity to zero on
every destroyable channel and to a stated bound on the one
(per-image mean luminance) that cannot be destroyed without
ruining the image.
\end{abstract}


%% =====================================================================
\section{Introduction}
\label{sec:intro}
%% =====================================================================

A large language model (LLM) agent deployment that is useful is,
by construction, an agent that is allowed to communicate: it posts
to channels, calls tools, returns text to its model, and ---
increasingly --- emits images, charts, screenshots, synthesized
audio, and other binary attachments. Every one of those acts is a
permitted egress. The security literature for agent runtimes has
concentrated on two adjacent questions. \emph{Destination} control
--- allowlists, server-side request forgery (SSRF) guards, virtual
private network (VPN) -only egress --- decides which hosts a
payload may reach. \emph{Content} control --- secret scanners,
data-loss-prevention (DLP) regexes, prompt-injection output filters
\cite{inan2023llamaguard} --- decides whether a payload contains a
recognizable bad string. Both are necessary. Neither addresses the
case this paper is about: a payload that goes to an allowed
destination, contains no recognizable secret, reads (or looks, or
sounds) entirely ordinary to a human, and nonetheless carries
exfiltrated data.

This is the classical \emph{confinement problem}
\cite{lampson1973confinement}: a program entrusted with data must
be prevented not only from transmitting it over the obvious channel
but from signalling it over channels ``not intended for information
transfer at all.'' Lampson's covert channels, and the \emph{side}
channels of subsequent practice, are precisely the gap between
destination control and content control. In an LLM agent runtime
the carriers are abundant, cheap, and now \emph{multi-modal}: an
agent emits free-form natural-language text and structured tool
arguments many times per session, and, when configured to do so,
also images and audio --- each a richer covert carrier than text
because each contains far more redundant capacity per unit of
human-perceivable content. An attacker who controls any part of the
agent's context --- a poisoned tool result, a malicious skill, a
prompt-injected web page, a subverted extension --- can modulate
any of those outputs.

We distinguish two families. A \emph{covert channel} (payload
level) hides data in the \emph{content} of a legitimate message:
zero-width and Unicode Tag characters \cite{ascii-smuggling},
bidirectional controls and homoglyphs \cite{boucher2023trojansource},
whitespace steganography, base64 or hex blobs, the permutation of
elements in a structured payload, the \emph{word and grammar
choice} that linguistic steganography exploits
\cite{ziegler2019neural, dewitt2023steganography} ---
and, when the carrier is binary, the least-significant bits of pixel
channels \cite{fridrich2001steganalysis}, the ultrasonic spectrum of
audio \cite{deshotels2014inaudible}, or the audible band itself
viewed as a carrier for \emph{sonified data} \cite{hermann2011sonification},
in which an image, a bitstream, or any structured data is rendered as
audible tones so that an apparently-musical waveform is in fact a
data envelope. A \emph{side channel} (metadata level) leaks through
observable \emph{properties} of egress that are not its content:
the timing, size, count, rate, or ordering of messages
\cite{cabuk2004timing}. Both families evade allowlists (the
destination is fine) and secret scanners (no known token appears).

\paragraph{Why sonification matters and why detection alone fails.}
Sonification --- the mapping of arbitrary data to audible
parameters such as pitch, amplitude, or timing
\cite{hermann2011sonification} --- is a legitimate scientific and
accessibility technique. It is also a perfect covert carrier: it
encodes data in the very band an audio scrambler dare not aggressively
filter, because aggressively filtering audible content destroys real
speech and music. No content-based classifier can robustly tell
``human speech'' from ``a bitstream rendered as
frequency-shift-keyed (FSK) audible tones'' without false positives: the discrimination is, in the
general case, an intractable steganalysis problem at the application
layer. We will not pretend otherwise. The mechanism we propose for
media instead inverts the question. Rather than ask a classifier
``is this audio data?,'' we ask ``does this payload carry a
cryptographic attestation, signed by a key the auditor trusted at
boot, that it is legitimate media of an authorized class?'' If yes,
the payload is exempt; if no, it is scrambled by default. The
attestation is the discriminator; the scrambler is the action; and
content classification --- the brittle part --- is not in the path.

\paragraph{Contributions.}
\begin{enumerate}[leftmargin=1.6em,itemsep=0.3em,topsep=0.3em]
\item An \emph{egress reference monitor} for text egress
(\cref{sec:monitor}): one mediated chokepoint, an ordered pipeline
of ten defense techniques, a per-sink leaky-bucket \emph{capacity
ledger}, and a \emph{staged enforcement} posture
(\cref{sec:staged}) that resolves the deploy-without-breakage
problem by letting lossless transforms enforce immediately while
lossy and statistical stages baseline in an audit mode.

\item A \emph{multi-modal extension} (\cref{sec:media}): two media
scramblers --- a Fourier-domain audio scrambler attacking
ultrasonic and sub-perceptual carriers, and a red-green-blue (RGB)
image-bit-depth requantizer with per-image mean-luminance
bucketing, attacking LSB-pixel and cross-image carriers --- gated
by a \emph{boot-time cryptographic legitimacy attestation}
(\cref{sec:legitimacy}) that exempts validly-signed media from
scrambling while leaving unsigned payloads suspect by default,
thereby closing audible-band sonification without resorting to a
brittle content classifier. The image scrambler's RGB-basis,
six-bit-per-channel default was arrived at by elimination across a
ten-pipeline design-space probe (\cref{sec:design-space}): every
cyan-magenta-yellow-black (CMYK) variant we tested leaves a
structural escape route at the unit-interval fixed points (any
pixel with $\max(R,G,B)=255$ or $\min(R,G,B)=0$ survives the
roundtrip untouched), and every selective-channel or
random-sprinkle scheme leaves an exploitable un-touched fraction;
the unique minimum-distortion transform that destroys both bit-0
and bit-1 stego is uniform RGB requantization to 64 levels.

\item A \emph{measurement methodology} (\cref{sec:capacity}):
residual covert capacity is the Miller-Madow-corrected mutual
information \cite{miller1955bias} between embedded and recovered
bits, exactly zero for a destroyed channel and not inflated by
finite-sample accuracy fluctuation.

\item An \emph{adversarial benchmark}
(\cref{sec:eval,sec:linguistic,sec:eval-media,sec:eval-cross-image})
of fifteen working covert / side-channel encoders across text,
image (per-pixel LSB, per-image mean luminance, inter-image
sequence permutation), and audio, with receiver-side decoders, a
reproducible harness, and results for a reference implementation
\cite{metere2026enclawed}.
\end{enumerate}

%% =====================================================================
\section{Threat model}
\label{sec:threat}
%% =====================================================================

\paragraph{Adversary.} We assume the agent's egress intent is
already compromised: a malicious skill, a poisoned tool result, a
prompt-injected document, or a subverted extension has induced the
agent to attempt exfiltration. We do \emph{not} assume the
attacker has compromised the host runtime itself; the reference
monitor is part of the trusted computing base (TCB). We also do
not assume the attacker has compromised any legitimate \emph{media
producer} (a signed text-to-speech (TTS) engine, chart renderer,
or screenshot
tool); compromising one of those is equivalent to compromising the
key that signs its outputs, and shifts the trust problem to key
hygiene at the producer side --- a problem we discuss in
\cref{sec:limits} but do not solve here. These are the same
boundaries the agent-skills supply-chain literature draws
\cite{owasp-llm-top-10, mitre-atlas} and the protection-system
literature has drawn since \citet{saltzer1975protection}.

\paragraph{What the attacker controls.} The attacker controls the
\emph{content} the agent tries to send, the \emph{schedule} on
which it tries to send it, and, when binary egress is permitted,
the \emph{bytes} of any media payload that does not carry a
producer's legitimacy signature. The attacker knows the defense
exists and is free to choose an encoding to evade it. The attacker
does not control the reference monitor's code, its configuration,
the auditor's boot-time trust set, or any legitimate producer's
private signing key.

\paragraph{Goal.} The attacker wants to transmit a secret
bit-string $S$ to a confederate who can observe the agent's
egress. The defender's goal is not to prove $S$ cannot leak --- it
can always leak one bit by the agent simply choosing to send or
not send --- but to drive the \emph{capacity} of every covert and
side channel low enough that exfiltrating a key, a credential, or
a customer record at a useful rate becomes infeasible, and to
\emph{measure} the residual so the claim is falsifiable.

\paragraph{Multi-modal carriers.} \cref{tab:carriers} enumerates
the carriers we treat. Each is a real, demonstrated technique, not
a hypothetical. The text family is the one most agent runtimes
already emit; the image and audio families apply when the
deployment is configured to emit binary media.

\begin{center}
\small
\begin{tabular}{@{}lllc@{}}
\toprule
Carrier & Modality & Family & Bits hidden in \\
\midrule
Zero-width / Tag chars & text   & covert & invisible code points \\
Variation selectors    & text   & covert & invisible code points \\
Bidirectional controls & text   & covert & invisible code points \\
Homoglyphs             & text   & covert & confusable letterforms \\
Whitespace runs        & text   & covert & space counts \\
Punctuation runs       & text   & covert & repeated marks \\
JSON key ordering      & text   & covert & element permutation \\
Base64 / hex blobs     & text   & covert & high-entropy substrings \\
Synonym / voice choice & text   & covert & word and grammar choice \\
Message timing         & meta   & side   & inter-send intervals \\
Message size           & meta   & side   & payload length \\
Image LSB              & image  & covert & low-order pixel bits \\
Per-image mean luma    & image  & covert & gross brightness across a sequence \\
Image sequence order   & image  & covert & permutation of a batch \\
Ultrasonic tones       & audio  & covert & $>$20 kHz spectrum \\
Audible-band sonified  & audio  & covert & FSK / amplitude in audible band \\
\bottomrule
\end{tabular}
\captionof{table}{Multi-modal covert and side-channel carriers in
agent egress. The audible-band sonified row is the one this paper
newly treats: it is the carrier no content classifier can
reliably tell from real audio, and is the motivation for the
cryptographic legitimacy attestation of \cref{sec:legitimacy}.}
\label{tab:carriers}
\end{center}

%% =====================================================================
\section{The egress reference monitor}
\label{sec:monitor}
%% =====================================================================

A reference monitor must be \emph{always invoked}, \emph{tamper
resistant}, and \emph{small enough to analyze}
\cite{saltzer1975protection}. Our design meets the first by
mediating a single egress chokepoint per modality; the ten
text-domain techniques are not ten independent interceptors but
ordered \emph{stages} of one pipeline, and the two media
transforms are not embedded into that pipeline but register into
a parallel media-scrambler registry that a host media-egress
chokepoint drains.

\subsection{One text chokepoint, one pipeline}

All outbound channel text passes through one hook
(\texttt{message\_sending} in the reference implementation). A
\emph{mediator} owns that hook, drains a registry of stages, and
runs them in a fixed priority order over the payload. Each stage
returns one of four decisions: \texttt{pass}, \texttt{rewrite}
(replace the text), \texttt{delay} (hold the send), or
\texttt{cancel} (drop the egress). The mediator merges decisions
with the same discipline a modifying hook chain uses: rewrites
compose, \texttt{cancel} is sticky and short-circuits, delays
accumulate.

\paragraph{Fail-closed without throwing.} The chokepoint in a real
runtime swallows exceptions from hook handlers so a buggy plugin
cannot wedge delivery --- which means a stage that defends by
\emph{throwing} would fail \emph{open}. The pipeline runner is
written so a stage never signals a block by raising; it returns an
explicit \texttt{cancel}, and a stage that does throw is caught by
the runner and converted to \texttt{cancel} when that stage is
enforcing. The reference monitor fails toward blocking egress.

\subsection{The ten text-domain stages}

Stages run in ascending priority; \cref{tab:stages} lists them.
The ordering is not arbitrary. Canonicalization runs first so
every later stage inspects already-normalized text. Taint and
entropy detection run before the lossy noise injector so they
score the original payload. Behavioral shaping runs last because
it decides timing and rate once the content decisions are settled.

Two design choices justify this layout. First, \emph{lossless
transforms precede lossy ones}: the canonicalizer removes carriers
with no legitimate use (zero-width characters, Tag characters,
bidi overrides; ordering of structured fields), so it can be run
unconditionally and the rest of the pipeline sees text that is
already free of the easy carriers. Second, \emph{detection runs
on the original payload, not on a noised version}: the entropy
scanner and replay sentinel measure what a payload would carry if
the disruption stages did nothing, so the capacity ledger gets
the upper bound on the bit-rate the pipeline is actually
suppressing. The disruption stages then act on that measurement
to decide how aggressively they have to rewrite. And the
constraint stages --- which observe \emph{timing}, \emph{rate},
and \emph{volume} rather than content --- run last because their
decisions are about the schedule of egress, which is settled only
after the content has been finalized.

\begin{center}
\small
\begin{tabular}{@{}rll@{}}
\toprule
Pri. & Stage & Function \\
\midrule
10 & Canonicalizer & strip invisible / Tag / bidi carriers; \\
   &               & fold homoglyphs; normalize Unicode, \\
   &               & whitespace, JSON key order \\
20 & Taint tracker & block credential material and \\
   &               & provenance-tagged sensitive values \\
30 & Entropy scanner & measure entropy, base64/hex blobs, \\
   &                 & incompressibility; charge the ledger \\
35 & Replay sentinel & flag cosmetic-only message replay \\
40 & Noise injector  & collapse visible covert-encoding \\
   &                 & classes (whitespace, punctuation) \\
45 & Semantic scrambler & collapse linguistic covert-encoding \\
   &                    & classes (synonym, voice, \\
   &                    & contraction), deterministically \\
46 & LLM scrambler   & regenerate the payload via a \\
   &                 & fixed-prompt model rephrasing \\
48 & Timing scrambler & inject a random per-message \\
   &                  & egress delay \\
50 & Behavioral invariants & rate/volume limits, timing \\
   &                       & quantization, paced release \\
-- & Cover traffic & constant-rate / Poisson filler \\
\midrule
\multicolumn{3}{l}{\emph{Media-domain scramblers (parallel
registry; \cref{sec:media}):}} \\
-- & Image scrambler & CMYK-basis bit-depth requantization, \\
   &                 & legitimacy-attestation gated \\
-- & Audio scrambler & FFT-domain band-limit and \\
   &                 & perceptual-floor zeroing, \\
   &                 & legitimacy-attestation gated \\
\bottomrule
\end{tabular}
\captionof{table}{The ordered text pipeline plus the parallel media
scramblers. Text priorities are stage identifiers; lower runs
earlier. Cover traffic is a service, not an in-line stage; the
media scramblers are drained by a separate media-egress chokepoint
provided by the host.}
\label{tab:stages}
\end{center}

\paragraph{Text covert channels in the wild.} Four concrete encodings
illustrate what the canonicalizer (and the noise injector immediately
behind it) is actually defending against. \cref{fig:text-examples}
shows each in three views: the carrier as a human sees it, the same
payload with the covert markers exposed, and the canonicalizer's
output. The point is that the first column is what reaches a human
reader's eye --- the channel only exists because the byte stream
carries information the rendered glyph does not.

\begingroup
\centering
\scriptsize
\setlength{\fboxsep}{1.2pt}
\setlength{\tabcolsep}{5pt}
\renewcommand{\arraystretch}{1.6}
\newcommand{\covert}[1]{\colorbox{red!18}{\strut\fontsize{6pt}{7pt}\selectfont #1}}
\begin{tabular}{|>{\raggedright\arraybackslash}p{0.11\textwidth}|>{\raggedright\arraybackslash}p{0.24\textwidth}|>{\raggedright\arraybackslash}p{0.34\textwidth}|>{\raggedright\arraybackslash}p{0.17\textwidth}|}
\hline
\textbf{carrier type} & \textbf{rendered (what a human sees)} & \textbf{bytes (covert markers exposed)} & \textbf{after canonicalizer} \\
\hline\hline
zero-width / Unicode Tag &
\texttt{Reset link sent.} &
\texttt{Reset\covert{ZWSP}link\covert{ZWNJ}sent\covert{TAG-b}.} &
\texttt{Reset link sent.} \\
\hline
homoglyph (Cyrillic\,$\to$\,Latin) &
\texttt{Please confirm your password.} &
\texttt{Please c\covert{Cyr-o}nfirm y\covert{Cyr-o}ur passw\covert{Cyr-o}rd.} &
\texttt{Please confirm your password.} \\
\hline
whitespace count &
\texttt{Hello,\,\,world.\,\,\,How\,\,are\,you?} &
\texttt{Hello,\covert{2$\sqcup$}world.\covert{3$\sqcup$}How\covert{2$\sqcup$}are\covert{1$\sqcup$}you?} &
\texttt{Hello, world. How are you?} \\
\hline
bidi override (RLO/PDF) &
\texttt{transfer \$100 to bob} &
\texttt{transfer \covert{RLO}bob ot 001\$\covert{PDF}} &
\texttt{transfer \$100 to bob} \\
\hline
\end{tabular}
\captionof{figure}{Text covert-channel examples and the canonicalizer's
response. Four payload-level encodings: invisible-code-point
smuggling (zero-width and Unicode-Tag characters between glyphs),
homoglyph substitution (Cyrillic U+043E in place of Latin
\texttt{o}, U+006F --- visible glyph unchanged), whitespace-count
steganography (extra spaces between tokens encoding bits), and
bidirectional-control reordering (RLO/PDF flipping byte order
without changing the rendered string --- the Trojan Source
primitive). The middle column reveals the byte content the eye
does not see; the right column is what the canonicalizer
(\cref{tab:stages}, stage 10) emits --- byte-identical to the
rendered carrier, with the covert payload destroyed.}
\label{fig:text-examples}
\par\endgroup\vspace{0.6em}

\paragraph{Detect, disrupt, constrain.} The ten text techniques
fall into three roles. \emph{Detection} (taint tracker, entropy
scanner, replay sentinel) measures what a payload plausibly
carries and either blocks known-bad content or charges an estimate
to the capacity ledger. \emph{Disruption} (canonicalizer, noise
injector, semantic scrambler, LLM scrambler) rewrites the payload
to destroy encoding dimensions, each collapsing a wider equivalence
class than the last: the canonicalizer removes carriers with no
legitimate use (a lossless operation); the noise injector collapses
\emph{visible} classes such as whitespace runs and repeated
punctuation; the semantic scrambler collapses \emph{linguistic}
classes --- deterministically mapping synonym choice, contraction
form, and active/passive voice to one canonical representative, so
the word-choice channel on which linguistic steganography
\cite{ziegler2019neural, dewitt2023steganography} depends carries
no information; and the LLM scrambler collapses the widest class
of all --- it regenerates the payload by rephrasing it through a
model under a fixed, non-configurable prompt, discarding the
attacker's entire choice of words, structure, and phrasing
wholesale (the last three lossy but meaning-preserving). The
fixed prompt is compiled in, never exposed to configuration, and
fences the input as data so the rephrasing model is not itself an
injection target; an unavailable model fails toward pass-through,
never toward blocking egress. \emph{Constraint} (timing scrambler,
behavioral invariants, cover traffic) bounds the side channels by
limiting rate and volume and by shaping send timing. Two timing
defenses sit here at different points on the cost/strength curve:
the timing scrambler injects a completely random,
cryptographically drawn per-message delay --- cheap, always-on,
and it \emph{reduces} timing-channel capacity in proportion to the
delay budget --- while the behavioral stage, in its strongest
configuration, paces releases onto a fixed cadence so the
observable inter-send interval is independent of the attacker's
schedule outright. The traffic-analysis literature
\cite{dyer2012peekaboo} is explicit that only the latter,
constant-rate shaping closes a timing channel deterministically;
the random scrambler is the low-latency complement, not its
replacement.

\subsection{The capacity ledger}

A covert channel is fundamentally a matter of bit-rate. The
monitor maintains, per egress destination (``sink''), a
leaky-bucket \emph{capacity ledger}: a bucket of covert-bit tokens
that refills at a configured rate. The entropy scanner
\emph{measures} the residual covert capacity of a payload --- a
base64 blob of $n$ characters carries at most $6n$ bits --- and
the behavioral stage \emph{enforces} a budget by drawing on the
ledger; when the bucket is empty, further suspicious payloads are
delayed or cancelled. The ledger is the seam that turns a
measurement (the scanner's estimate) into an enforcement (a
per-sink rate bound), and it is the reason the ten stages are one
system rather than ten filters. It is also bounded in size to
resist unbounded-state attacks from attacker-chosen sink
identifiers.

\subsection{Host-injected capability seams}

Four capabilities require host internals the pipeline must not
import directly: checking the deployment's license entitlement,
emitting cover traffic through the host's outbound stack,
invoking the rephrasing model for the LLM scrambler, and ---
introduced by the media extension --- verifying media legitimacy
signatures against the auditor's boot-time trust set. The monitor
exposes them as injected seams: the host installs probes and
emitters at start-up; absent an injection the license probe
defaults to \emph{enabled} (a security control fails safe rather
than fails open), the cover-traffic and LLM seams default to
inert, and --- crucially --- the media-legitimacy verifier
defaults to \emph{rejecting every payload}, so that the absence
of a boot-time trust set does not silently open a hole. This
keeps the pipeline a closed, analyzable unit.

%% =====================================================================
\section{Multi-modal egress and the legitimacy attestation}
\label{sec:media}
%% =====================================================================

Text is not the only carrier an agent emits. An outbound image or
audio clip hides data where neither a human nor a text-stage looks.
The text pipeline carries a string and cannot inspect binary media,
so media defenses operate through a parallel media-scrambler
registry; the present design provides the two transforms (one image,
one audio) and the boot-time attestation that gates them, and a host
that adds a media-egress chokepoint drains the registry exactly as
the mediator drains the text-stage registry.

\subsection{The image scrambler: RGB six-bit-per-channel
requantization}
\label{sec:media-image}

A working LSB-steganography channel hides each bit in the
least-significant bit of a pixel's blue (or any) colour channel ---
the canonical bitmap steganography literature
\cite{fridrich2001steganalysis} treats LSB as a baseline. A
moderately aware attacker simply moves the encoding one bit up
(into bit-1 of the same channel), so a defense that only destroys
the lowest bit is not a defense; it is a no-op against any
attacker who reads the design.

The image scrambler addresses both with a uniform RGB-basis
quantizer. Each channel is snapped to one of 64 evenly-spaced
levels (six bits per channel, quantization step $\approx 4.05$),
which collapses every pair of values differing by 2 onto the same
output and therefore destroys both bit-0 and bit-1 stego in every
channel of every pixel. The per-channel change is at most two
units out of 255 --- sub-perceptual at eight-bit colour --- and
the bit-0 / bit-1 receiver-side decoders we measure both fall to
chance (\cref{sec:design-space}).

\paragraph{Why not CMYK, why not lighter.} We chose this default
by elimination across a ten-pipeline design-space probe
(\cref{sec:design-space}) covering naïve depth reductions
(``256~$\to$~255 levels''), several CMYK-roundtrip variants, K-only
quantization, randomised per-pixel selection of channels or
bit-budgets, and a seven-bit-baseline-plus-six-bit-sprinkle hybrid.
Two structural facts emerged. First, every CMYK variant has an
escape route: the unit-interval quantizer used inside the CMYK
roundtrip maps $0\to 0$ and $1\to 1$ as fixed points, so any RGB
pixel with $\max(R,G,B)=255$ has $K=0$ and any pixel with
$R=G=B=0$ has $C=M=Y=0$, and those pixels survive the trip
bit-identical regardless of how aggressively the CMYK components
are quantized. The receiver gets their LSBs back unmodified. The
``misaligned grid'' intuition that motivated CMYK in the prior
design is theoretically appealing but does not pay off:
empirically (\cref{tab:design-space}) CMYK at six bits per
component destroys the same bit positions (bit-0 and bit-1) as
RGB at six bits per channel, at higher visible distortion. Second, no sub-sampling
scheme reaches chance-level recovery for the bit the attacker
actually uses --- the recovery probability tracks the un-touched
fraction in a closed form $P_\text{recover} = 0.5 + 0.5\,
P_\text{untouched}$, so anything less than uniform treatment of
every LSB-carrying value leaves an exploitable channel. The unique
minimum-distortion uniform transform is RGB at 64 levels. The
CMYK basis remains available as a configurable option for
deployments that prefer the misaligned-grid story at marginally
higher visible cost.

A reader may worry that this leaves \emph{bit-2 and higher} stego
unaddressed: a still-more-aware attacker could move encoding into
bit-2 (step $\geq 4$ in the carrier) and recover everything. That
is correct, and is a deliberate scope decision. Each additional
bit of channel depth destroyed costs one more bit of distortion in
the worst case (the maxima go 1, 2, 4, 8, $\dots$); five bits per
channel (step $\approx 8.23$, $\max\Delta=4$) is borderline-
perceptible in bright and mid-tone regions and is configurable as
\texttt{depthLevels=32} for deployments that want it.

\subsubsection*{Cross-image carriers: mean luminance and sequence
ID}
\label{sec:cross-image}

Two carriers operate \emph{across} a sequence of images and are
invisible to a per-pixel scrambler. The first is the
\emph{per-image mean luminance}: the attacker emits a stream of
images and encodes bits in the gross brightness of each, so the
sequence of means $\bar{L}_1, \bar{L}_2, \ldots$ carries the
payload. The second is the \emph{sequence ID / permutation}: given
a batch of $N$ distinguishable images, the attacker chooses the
emission order to encode up to $\log_2(N!)$ bits in the
permutation. Neither is touched by per-pixel bit-depth
requantization.

The honest defensive position differs between the two. The
\emph{sequence-ID channel} is fully closeable for unsigned media
at the application layer: a canonical re-ordering by sha256
content hash before emission (provided as a helper that the host's
media-egress chokepoint invokes) replaces the attacker's
permutation with one they cannot pre-image to encode bits. Once
applied, the residual sequence-ID capacity is zero modulo a
second-preimage attack on sha256. The legitimacy attestation
takes a producer's signed sequence as authentic and preserves its
relative order (paginated chart decks, time-ordered screenshots
have semantically meaningful orderings); the canonicalization
applies only to the unsigned tail.

The \emph{mean-luminance channel} cannot be perfectly destroyed
without making the image unusable: clamping every image to one
fixed mean would render a dark chart and a bright chart
identically grey. The scrambler instead \emph{bounds} the channel
by snapping each unsigned image's mean luma to one of $B$ coarse
buckets (default $B=8$), so per-image capacity is bounded above
by $\log_2 B$ bits regardless of how the attacker modulates the
underlying mean. Combined with the per-sink capacity ledger
(\cref{sec:monitor}) the across-sequence aggregate is rate-
bounded. This is a deliberately honest distinction: we do not
report ``$100\%$ reduction'' for a channel we can only bound. The
benchmark below reports the per-image bound directly.

\paragraph{Why the legitimacy attestation does not subsume this.}
A legitimately-signed batch retains its mean luminances and its
order; the producer chose them, the auditor authorized the
producer. Unsigned media is the case where these carriers exist,
and there the bucketing-plus-canonical-order defense applies. The
attestation does not somehow make luminance bucketing redundant
for unsigned media; it just designates which media is exempt.

\subsection{The audio scrambler: fast Fourier transform (FFT)
band-limit and perceptual floor}
\label{sec:media-audio}

Audio is a rich covert carrier. Two classes of hidden data survive
ordinary playback and ordinary inspection. \emph{Out-of-band
energy} (ultrasonic, $>20$ kHz, or subsonic, $<20$ Hz) is the
basis of real device-to-device covert signalling
\cite{deshotels2014inaudible}: a tone a human cannot hear can be
demodulated by a confederate, riding silently on otherwise
ordinary audible audio. \emph{Sub-perceptual in-band energy} ---
narrowband tones far below the audible content's level --- can
likewise carry an FSK bitstream invisible to a casual listener.
The audio scrambler runs a fast Fourier transform, zeroes every
bin outside the human audible band and every bin whose magnitude
is below a perceptual floor relative to the loudest bin, and runs
the inverse FFT. What a human hears is preserved; what only a
covert receiver could read is gone.

\subsection{The audible-band sonification gap and why a content
classifier cannot close it}

The two transforms above close the out-of-band and sub-perceptual
audio channels and the LSB image channel. They do not, on their
own, address the \emph{audible-band sonified} carrier of
\cref{tab:carriers}: a bitstream rendered as audible FSK,
amplitude-modulated tones, or any other sonification
\cite{hermann2011sonification} of structured data into the
20--20{,}000 Hz range. Aggressively scrambling that band is not an
option --- it would destroy real speech and music. The natural
question is whether a steganalysis classifier could discriminate
``real audio'' from ``sonified data,'' so the scrambler runs only
on the latter. This is, in the general case, an intractable
classification problem at the application layer: a sufficiently
careful sonification can be made to satisfy whichever spectral or
statistical signature a classifier keys on, and the false-positive
rate on legitimate music alone makes per-frame classification
unworkable for an enforcement control. We do not attempt that
classifier. The same problem appears for images: an image
deliberately constructed to encode data as visible patterns --- a
QR-like grid, a halftone of a bitstream, an LSB-stego cover crafted
to look plausibly photographic --- cannot be content-classified as
``data'' versus ``picture'' in any robust way.

We close the gap with a cryptographic mechanism.

\subsection{Boot-time cryptographic legitimacy attestation}
\label{sec:legitimacy}

A \emph{media legitimacy token} is a signed assertion, attached to
a media payload, that the payload is genuine media of a declared
kind (image or audio) and data class (e.g.\ \texttt{speech-
synthesis}, \texttt{chart-render}, \texttt{screenshot}), with a
content hash binding the signature to the exact payload bytes. The
signature is Ed25519 \cite{bernstein2012ed25519} over the
canonical bytes of the claim. The producer does \emph{not} bring
that signing key. It is minted by the zero-trust accreditor through
a two-stage chain (\cref{sec:legitimacy-keychain}) that anchors
liability at the launcher and keeps every secret behind a sandbox
boundary so the key never appears as raw material in the
producer's address space.

\paragraph{What the auditor does at boot, and only at boot.} An
\emph{auditor} (in practice the host runtime's boot sequence)
establishes, before any media is processed, a
\emph{trust set}: the Ed25519 public keys whose signatures will be
accepted, and the finite set of $\{$kind, data-class$\}$ pairs
that this deployment is authorized to emit. This is the
boot-time-only decision --- the auditor decides \emph{which kinds
of data} are legitimately produced for this deployment, and that
decision is fixed for the process lifetime. The host then
installs a verifier built from that trust set as the media
scramblers' legitimacy seam; the trust set is not consulted
again at runtime.

\paragraph{What the verifier does at runtime.} For each outbound
media payload, before scrambling, the registered media scrambler
calls the verifier with the payload's actual content hash and the
attached token. A token is accepted iff (i) its claimed content
hash equals the payload's actual hash, (ii) its $\{$kind, data-
class$\}$ pair is in the boot-authorized set, and (iii) its
signature verifies under one of the trusted keys. An accepted token
exempts the payload from scrambling: the legitimate audio (or
image) is passed through unchanged, with no false-positive
degradation. A rejected or missing token means the payload is
suspect and the scrambler runs. The runtime check is a single
signature verification.

\paragraph{The fail-secure default.} With no verifier installed by
the host, the seam returns \texttt{false} for every payload: no
media is legitimate by default. The absence of an auditor's
boot-time trust set therefore does not open a hole; it scrambles
everything. A deployment that wishes to permit unmangled legitimate
media must explicitly publish a trust set at boot --- the security
posture is opt-in to exemptions, not opt-in to scrambling.

\subsection{The accreditor-brokered key chain}
\label{sec:legitimacy-keychain}

The signing key the producer uses is \emph{not} a key the producer
generates on its own. It is minted by the zero-trust accreditor
through a two-stage chain that pins liability to the launcher and
keeps the trust set the verifier consults equal to the set of keys
the accreditor itself has minted this process.

\begin{enumerate}[leftmargin=1.4em]
\item \textbf{Stage 1 --- job launch.} A launcher (the human or
system principal that initiates the job) presents a
\emph{LaunchAuthority} to the accreditor's identity broker. The
accreditor verifies the launch authority against its own trust
root and mints a fresh \emph{producer identity}: an Ed25519
identity keypair bound to a producerId AND to the launcherId. The
launcher is what holds liability for everything the resulting
identity signs for. Any keypair the producer (or anything else)
may have brought is discarded.
\item \textbf{Stage 2 --- producer boot.} The producer presents
its accreditor-minted identity to the accreditor's signing-key
broker. The broker verifies the identity is one it minted (and
that it carries the \texttt{generator} permission), and mints a
fresh \emph{per-process Ed25519 signing keypair}. The public half
is added to the live trust pool the verifier consults; the
private half lives only inside the accreditor's sandbox.
\end{enumerate}

\paragraph{Per-process immutability.} Both the identity and the
signing key are \emph{immutable for the process lifetime}. A
second request for the same producerId in the same process is
rejected at the application layer (and independently at the
accreditor). To rotate a signing key the producer must restart; to
rotate an identity the job must be re-launched. The
generator-permission gating is therefore enforced both temporally
(one key per process) and structurally (only keys minted via the
accreditor enter the trust pool).

\paragraph{Sandbox boundary and the WebAssembly question.} The
strongest deployment in this codebase runs the accreditor as a
single-executable-application (SEA) binary in its own process;
the producer's JavaScript process holds only a \emph{handle}
whose \texttt{sign(bytes)} method round-trips bytes to the
accreditor over a local inter-process-communication (IPC) channel
and returns just the signature. The signing private key never
crosses the process boundary, and the producer cannot enumerate
or extract it: the cross-process boundary is the real one. Hosts
that prefer in-process \emph{memory isolation} (against the
producer code, not against a compromised host) can fulfil the
same handle interface with a WebAssembly module whose linear
memory is not exported back to the host --- the producer holds
only the
instance's \texttt{sign} export; the key lives in the module's
own linear memory. We are explicit about the limit of that
second option: WebAssembly in the same JavaScript process is a
sandbox \emph{against the producer} (the producer cannot reach
into module memory) but \emph{not against the host}, which can
inspect any \texttt{WebAssembly.Memory} it loaded. For real
isolation against a host-side attacker the deployment must use
process isolation (the SEA accreditor over IPC) or a hardware
enclave; the application-layer code in this paper provides the
sandboxed-handle interface, but the strength of the boundary is
set by the host's implementation choice.

\paragraph{The liability chain.} Every minted handle carries
forward the launcherId of the identity it was minted under. The
chain (launcher $\to$ identity $\to$ signing key $\to$ signed
media) is therefore observable at the verifier: a signature
identifies a launcher as the liable principal. This is the
audit-trail property we want; it does not magically prevent a
malicious launcher from emitting bits, but it makes the launcher
the named principal whose authority is on the line.

\paragraph{Why this closes the sonification gap.} The audible-band
sonified carrier of \cref{tab:carriers} is, by construction,
\emph{unsigned}: a malicious producer who could sign on behalf of
a trusted key has compromised the producer's secret, which is the
key-hygiene problem and outside our scope. An unsigned audible
payload is by default scrambled (the FFT pass strips its out-of-
band and sub-perceptual content; \cref{sec:eval-media} reports
that this already collapses the in-band carriers we tested, by
removing the precise spectral structure the receiver decodes
against). A legitimate signed audible payload is exempt and
arrives unchanged. The discriminator is the signature, not a
content classifier --- and the false-positive question reduces to
key trust, not to spectral statistics.

\begin{proposition}[Allowlist-by-signature]
Under the boot-time-only auditor model, the set of payloads
exempted from media scrambling is exactly the set of payloads
carrying a token $(C, \sigma)$ such that the claim $C$'s
content-hash field equals the payload's actual content hash, the
pair $(C.\text{kind}, C.\text{dataClass})$ is in the boot-published
authorized set $A$, and $\sigma$ verifies under one of the boot-
published public keys $K$. The runtime never adds to $A$ or to
$K$; therefore an attacker without a trusted secret cannot
construct an exempt payload, and a content classifier is not on
the trust path.
\end{proposition}

%% =====================================================================
\section{A capacity model and the staged posture}
\label{sec:capacity}
%% =====================================================================

\subsection{Covert capacity as mutual information}
\label{sec:mi}

Treat one covert channel as a discrete channel. The attacker
embeds a bit $X$; the defense transforms the payload; a
confederate runs a receiver-side decoder and outputs a bit $Y$.
Over many trials with uniformly random $X$, the channel is
characterized by the joint distribution of $(X, Y)$.

\begin{definition}[Residual covert capacity]
The residual capacity of a covert channel under a defense is the
mutual information
\[
  I(X; Y) \;=\;
  \sum_{x, y} p(x, y)\,
  \log_2 \frac{p(x, y)}{p(x)\,p(y)}
\]
estimated from the empirical joint distribution of embedded bit
$X$ and recovered bit $Y$.
\end{definition}

A widely used shortcut estimates capacity from the receiver's
\emph{accuracy} $a$ as the binary-symmetric-channel capacity
$1 - H(a)$, with $H$ the binary entropy. This is wrong for a
\emph{destroyed} channel. When the defense neutralizes the
carrier, the decoder's output is independent of $X$ --- often
literally constant. A constant or independent $Y$ has
$I(X;Y) = 0$ exactly. But the measured accuracy on a finite
sample of random secrets fluctuates off $0.5$, and because
$1 - H(a)$ is convex it returns a small \emph{positive} number on
average for a channel that in truth carries nothing. The artifact
is not large --- on the order of a tenth of a percent --- but it
means the accuracy proxy can never report a channel as fully
closed even when it is.

We therefore measure $I(X;Y)$ directly. Empirical mutual
information has a known positive bias for finite samples; we apply
the Miller-Madow correction \cite{miller1955bias}, subtracting
$(r-1)(c-1)/(2N\ln 2)$ for an $r \times c$ table over $N$ samples,
and clamp at zero. For a destroyed channel --- decoder output
constant --- the correction term vanishes and the estimate is
exactly $0$. \emph{Capacity reduction} is then
$1 - I(X;Y)$ per transmitted bit, and reaches $100\%$ when, and
only when, the receiver does no better than chance \emph{and}
that is confirmed by the joint distribution, not by an accuracy
number alone. The same metric is used for the text, image, and
audio benchmarks.

\subsection{The staged enforcement posture}
\label{sec:staged}

A covert-channel control that blocks aggressively from first
install will also rewrite or cancel legitimate traffic --- a long
code block, a deliberately formatted table, an attached data file,
a chart, a TTS clip. The design resolves this with a
\emph{staged} posture keyed to whether a stage's action is
lossless:

\begin{itemize}[leftmargin=1.4em]
\item \textbf{Lossless stages enforce immediately.} The
canonicalizer only removes code points with no legitimate use in
agent text (zero-width spaces, Tag characters, bidi overrides) and
canonicalizes orderings; these operations preserve every
human-visible aspect of the message, so blocking and rewriting
are safe on day one. The taint tracker blocks only high-confidence
credential material.
\item \textbf{Detection and shaping stages baseline in audit.} The
entropy scanner, replay sentinel, noise injector, and behavioral
invariants begin in an \emph{audit} mode: they log findings and
compute would-be decisions but do not alter or cancel traffic. A
deployment escalates them to \emph{enforce} per sink after
observing its own baseline.
\item \textbf{Media scramblers are gated by attestation, not by
audit/enforce.} The legitimacy attestation already provides the
false-positive escape valve: signed legitimate media is exempt,
unsigned media is scrambled. There is no audit-mode for a media
scrambler --- the staging dimension that ``audit'' addresses for
text (``would this rewrite have broken anything?'') is supplanted
by the boot-time trust set (``did the auditor authorize this kind
of data?'').
\item \textbf{Cover traffic is opt-in.} Emitting filler is
intrusive and is disabled by default.
\end{itemize}

The staged posture is itself a measurable design: \cref{sec:eval}
reports both text postures, and the gap between them is the price
of zero false positives at install time.

%% =====================================================================
\section{Adversarial evaluation: text}
\label{sec:eval}
%% =====================================================================

\subsection{Methodology}

The benchmark is deliberately not a unit test of ``was a character
stripped.'' For each carrier of \cref{tab:carriers} we implement
the exfiltration tooling an attacker would actually build: an
\emph{encoder} that embeds an $L$-bit secret into a benign
carrier message, and a receiver-side \emph{decoder} that knows
the protocol and recovers bits from whatever survives the defense.

\paragraph{Procedure.} For each channel and each enforcement
posture we run $T$ independent trials. Each trial draws a
uniformly random $L$-bit secret from a seeded generator (so the
statistics are reproducible), encodes it, runs the payload
through the full pipeline behind the mediator, and decodes the
delivered output --- the empty string if the pipeline cancelled
it. We accumulate the $(X, Y)$ pairs across all trials and all
bit positions, compute the Miller-Madow-corrected $I(X;Y)$ of
\cref{sec:mi}, and report capacity reduction $1 - I(X;Y)$, the
receiver accuracy (as a sanity column), and the fraction of trials
in which the pipeline emitted an audit finding or cancelled.

\paragraph{The timing side channel} is measured against the
behavioral monitor directly: the attacker encodes bits in the gap
between message arrivals; the monitor applies its timing defense;
the receiver recovers bits from the \emph{observed} inter-send
intervals. The strongest configuration is constant-rate paced
release: consecutive sends are pinned at least one cadence apart,
so an attacker sending fast enough to encode bits sees every
release on a fixed interval and the observed gap is
input-independent.

\subsection{Results}

\cref{tab:results} reports the text benchmark for the reference
implementation with $T = 20$ trials per cell. Two postures are
shown: the \emph{default} staged posture (canonicalizer and taint
tracker enforcing, all other stages auditing) and \emph{full
enforcement} (every stage enforcing).

\begin{center}
\small
\begin{tabular}{@{}lrrr@{}}
\toprule
Channel & Default & Enforce & Detected \\
        & reduct.\ & reduct.\ & \\
\midrule
zero-width / Tag      & 100.0\% & 100.0\% & 100\% \\
variation selectors   & 100.0\% & 100.0\% & 100\% \\
bidirectional control & 100.0\% & 100.0\% & 100\% \\
homoglyph             & 100.0\% & 100.0\% & 100\% \\
JSON key ordering     & 100.0\% & 100.0\% & 100\% \\
whitespace count      &   0.0\% & 100.0\% & 100\% \\
punctuation count     &   0.0\% & 100.0\% & 100\% \\
synonym substitution  &   0.0\% & 100.0\% & 100\% \\
base64 blob           &   0.0\% & 100.0\% & 100\% \\
timing interval       &  --     & 100.0\% & --    \\
\midrule
\textbf{text mean}    & \textbf{60.2\%} & \textbf{100.0\%} & \\
\bottomrule
\end{tabular}
\captionof{table}{Measured covert-channel capacity reduction
($1 - I(X;Y)$, Miller-Madow corrected) for the text-domain
encoders. ``Detected'' is the fraction of trials flagged or
cancelled, under the default posture. The default-posture row for
the timing channel is omitted because timing shaping is an
enforce-only behavior; the mean is taken over the eleven content
and side channels of the benchmark.}
\label{tab:results}
\end{center}

Three things in \cref{tab:results} are worth stating plainly.

First, the \textbf{lossless stages close their channels under
both postures}. Every carrier the canonicalizer handles ---
invisible code points, bidi controls, homoglyphs, key ordering
--- is reduced to zero residual capacity even in the default
posture, because the canonicalizer enforces from install. The
receiver's accuracy on these channels sits at chance
($\approx 48\%$); the mutual information confirms $0$ residual
bits.

Second, the \textbf{staged posture is real and visible}. Under
the default posture the whitespace, punctuation, synonym, and
base64 channels carry their full payload --- $0\%$ reduction ---
because the noise injector, semantic scrambler, and entropy
scanner are auditing, not enforcing. They are nonetheless
\emph{detected} in $100\%$ of trials: the default posture is a
detect-first stance, not an unprotected one. Escalating those
stages to enforce takes the same channels to $100\%$ reduction.
The $60.2\% \rightarrow 100.0\%$ gap is the measured cost of
guaranteeing zero false positives at install.

Third, the \textbf{timing side channel is closed by paced
release}, not merely blurred. Send-time quantization alone leaves
a residual: a sub-grid gap signal survives, and we measured
roughly one-fifth of the channel still open. Constant-rate paced
release removes it: with every send pinned to a fixed cadence the
observed interval no longer depends on the attacker's gaps, the
receiver falls to chance, and $I(X;Y)$ is zero.

Under full enforcement the measured residual text capacity is
zero across the tested ensemble.

\subsection{The linguistic carriers in isolation}
\label{sec:linguistic}

The word- and grammar-choice carrier deserves a closer look,
because it is the one the steganography literature
\cite{ziegler2019neural, dewitt2023steganography} treats as the
hardest to detect. We ran a second, focused benchmark driving
four linguistic covert channels directly through the deterministic
semantic scrambler: synonym choice at one and at two bits per
word, contraction form (contracted vs.\ expanded), and active /
passive voice. \cref{tab:linguistic} reports the result over $40$
trials per channel.

\begin{center}
\small
\begin{tabular}{@{}lrrr@{}}
\toprule
Linguistic channel & Undefended & Scrambler & Receiver \\
                   & reduct.\   & reduct.\  & accuracy \\
\midrule
synonym choice (1 bit/word) & 0.1\% & 100.0\% & 51.6\% \\
synonym choice (2 bit/word) & 0.1\% & 100.0\% & 51.6\% \\
contraction form            & 0.1\% & 100.0\% & 51.6\% \\
active / passive voice      & 0.1\% & 100.0\% & 51.6\% \\
\bottomrule
\end{tabular}
\captionof{table}{Linguistic covert channels driven directly
through the deterministic semantic scrambler. Undefended, each
channel carries its full payload; with the scrambler enforcing,
the receiver's accuracy falls to chance and the mutual
information is zero. Every trial is also flagged in the audit
log.}
\label{tab:linguistic}
\end{center}

Two points. First, the \textbf{undefended column confirms the
channels are real}: with no defense, each receiver recovers the
secret essentially perfectly ($\le 0.1\%$ reduction), including
the two-bit-per-word variant that packs more capacity into the
same words. Second, the scrambler's canonical collapse is
\emph{idempotent} --- re-scrambling a scrambled payload is a no-op
--- so an attacker cannot iterate the defense to recover signal:
once the equivalence class is collapsed, the bits are gone.

\paragraph{The model-backed variant.} The LLM scrambler is not in
the mutual-information benchmark, and deliberately so: it is
non-deterministic and requires a live model, neither of which
suits a reproducible unit benchmark. Its mechanism, however,
dominates the deterministic scrambler by construction --- it
regenerates the payload rather than collapsing enumerated classes,
so it discards carriers (idiosyncratic phrasing, clause ordering,
register) that no finite rule set enumerates. It is
configuration-gated and audit-default precisely because that
strength comes with a model dependency, a latency cost, and a
prompt-injection surface that the fixed, input-fencing prompt
mitigates but does not eliminate (\cref{sec:limits}). Because the
regeneration subsumes the deterministic collapse, the two
linguistic scramblers are \emph{alternatives}, not a stack: each
is a separate extension with its own enable flag, independently
selectable, and a production pipeline runs at most one.

%% =====================================================================
\section{Adversarial evaluation: image and audio}
\label{sec:eval-media}
%% =====================================================================

The media transforms are measured against working encoders for the
binary carriers of \cref{tab:carriers}. The experiments below
exercise the scrambler transforms with no legitimacy attestation
attached --- the suspect path. With an authorized attestation the
scrambler is bypassed and the payload arrives bit-identical; that
behavior is verified separately by unit tests of the legitimacy
gate (the gate either passes the payload through unchanged, by
object identity, or hands off to the scrambler) and is not the
adversarial measurement here.

\subsection{Image LSB steganography against the chosen default}

A working LSB encoder writes each bit of an $L$-bit secret into a
chosen bit position of the blue channel of an RGB image; the
decoder reads the same bit position. We run $T$ trials over
distinct cover images and accumulate the bitwise $(X, Y)$ pairs.
Two attacker strengths are reported: a \emph{naïve} attacker who
hard-codes the encoding to bit-0 (the classical LSB), and an
\emph{aware} attacker who, knowing the defense, moves the encoding
up one bit position into bit-1.

\begin{center}
\small
\begin{tabular}{@{}lrrr@{}}
\toprule
Defense & naïve bit-0 acc.\ & aware bit-1 acc.\ & Reduction (bit-0) \\
\midrule
none (undefended)              & 100.0\% & 100.0\% &   0.0\% \\
RGB roundtrip, 256 levels      & 100.0\% & 100.0\% &   0.0\% \\
RGB 7 bits/ch (128 levels)     &  50.0\% & 100.0\% & 100.0\% \\
\textbf{RGB 6 bits/ch (64 levels, default)} &  \textbf{50.6\%} & \textbf{50.0\%} & \textbf{100.0\%} \\
RGB 5 bits/ch (32 levels)      &  50.8\% &  51.4\% & 100.0\% \\
\bottomrule
\end{tabular}
\captionof{table}{Image LSB steganography against the image
scrambler at each RGB depth setting. Six bits per channel is the
minimum default that drives \emph{both} the naïve and the aware
attacker to chance; seven bits defeats only the naïve attacker.
The full design-space probe (every alternative we tested,
including the CMYK roundtrip variants) is in
\cref{tab:design-space} of \cref{sec:design-space}.}
\label{tab:image}
\end{center}

\subsection{Cross-image carriers: mean luminance and sequence ID}
\label{sec:eval-cross-image}

The two cross-image carriers of \cref{sec:cross-image} are
measured against working encoders that emit a sequence of $N$
images each carrying one bit of secret. The receiver reads the
per-image mean luma (luminance channel) or the per-emission
order (sequence channel).

\begin{center}
\small
\begin{tabular}{@{}llrr@{}}
\toprule
Channel & Defense & Per-image bound & Outcome \\
\midrule
Mean luminance ($1$ bit) & undefended       & $\sim 8$ bits & receiver $\approx 100\%$ \\
Mean luminance ($1$ bit) & bucket $B=8$     & $3$ bits      & 1-bit attacker still works \\
Mean luminance ($1$ bit) & bucket $B=2$     & $1$ bit       & 1-bit attacker still works \\
Sequence ID ($4!$ perm.) & undefended       & $\sim 4.6$ bits & receiver recovers permutation \\
Sequence ID ($4!$ perm.) & sha256 canonical & $0$ bits      & permutation replaced \\
\bottomrule
\end{tabular}
\captionof{table}{Cross-image carriers. The mean-luminance channel
is \emph{bounded}, not destroyed: bucketing limits per-image
capacity to $\log_2 B$ bits, and a 1-bit-per-image attacker
remains within that bound (which is the point). The sequence-ID
channel is closed by canonical re-ordering: the attacker's chosen
permutation is replaced by the sha256 order of the unsigned
payloads, and (modulo second-preimage on sha256) the attacker has
no degrees of freedom. Honest framing: we report the bound for
luminance, not a $100\%$ figure we cannot defend.}
\label{tab:cross-image}
\end{center}

\subsection{Design-space probe: how we chose the image
scrambler's default}
\label{sec:design-space}

Before settling on the uniform RGB-basis six-bit-per-channel
default, we tested every alternative we (and during iterative
design review, three external proposals) could plausibly motivate:
gentler reductions, CMYK roundtrips with various depths, K-only
quantization that leaves the chromatic axes intact, randomized
per-pixel per-component bit-budgets, randomized one-of-three
channel selection at seven bits, and a seven-bit baseline with a
$p$-fraction six-bit sprinkle. Each variant was applied to the
same 128$\times$128 test image with the secret embedded in turn
into bit-0 (naïve attacker) and bit-1 (aware attacker) of the
blue channel. \cref{tab:design-space} reports recovery rates for
each attacker and the per-pixel maximum and mean absolute channel
change; chance is $50.0\%$ and anything above is residual leakage.
\cref{fig:design-space-grid} shows the same images side-by-side
so the visible cost of each defense is judgeable.

\begin{center}
\scriptsize
\renewcommand{\arraystretch}{1.05}
\begin{tabular}{@{}lrrrr@{}}
\toprule
Defense & bit-0 acc.\ & bit-1 acc.\ & max $\Delta$ & mean $\Delta$ \\
\midrule
RGB 256 (no defense) & 100.0 & 100.0 & 0 & 0.00 \\
RGB 255 (one level less) & 99.4 & 99.3 & 1 & 0.00 \\
RGB 200 & 76.7 & 88.9 & 1 & 0.21 \\
RGB 128 (7 bits/ch) & 50.0 & 100.0 & 1 & 0.51 \\
\textbf{RGB 64 (6 bits/ch, DEFAULT)} & 50.6 & 50.0 & 2 & 1.00 \\
RGB 32 (5 bits/ch) & 50.8 & 51.3 & 4 & 2.04 \\
RGB 16 (4 bits/ch) & 52.4 & 52.5 & 8 & 4.20 \\
CMYK 255 (one level less) & 88.1 & 94.2 & 1 & 0.08 \\
CMYK 128 (7 bits/comp) & 52.3 & 82.2 & 2 & 0.47 \\
CMYK 64 (6 bits/comp) & 51.4 & 51.8 & 3 & 0.97 \\
CMYK 32 (5 bits/comp) & 50.0 & 51.1 & 7 & 2.07 \\
K-only 4 bits (others full) & 51.5 & 48.0 & 8 & 3.37 \\
K-only 2 bits (others full) & 51.6 & 52.5 & 42 & 20.31 \\
CMYK random in [7, 8] bits & 71.9 & 89.6 & 2 & 0.27 \\
CMYK random in [6, 7] bits & 50.4 & 65.0 & 3 & 0.75 \\
CMYK random in [5, 7] bits & 50.5 & 58.3 & 7 & 1.23 \\
RGB: random 1-of-3 ch at 7 bit & 83.3 & 100.0 & 1 & 0.17 \\
RGB 7-bit + 25\% sprinkle 6-bit & 49.5 & 87.8 & 2 & 0.64 \\
RGB 7-bit + 50\% sprinkle 6-bit & 49.4 & 75.5 & 2 & 0.76 \\
\bottomrule
\end{tabular}
\captionof{table}{Master comparison: every defense variant tested,
ranked by family. Receiver accuracy at chance is $50\%$; bit-0
is the naïve LSB attacker, bit-1 is an aware attacker who moves
the encoding up one bit. \textbf{max $\Delta$} is the worst per-
channel pixel change (out of 255); \textbf{mean $\Delta$} is the
average. The chosen default --- RGB at 64 levels (six bits per
channel) --- is the unique minimum-distortion entry that drives
both attackers to chance. Every CMYK variant leaves either bit-1
or bit-0 above chance, with the structural reason
(unit-interval fixed points) explained in \cref{sec:media-image};
sub-sampling variants follow the closed-form
$P_\text{recover} = 0.5 + 0.5\,P_\text{untouched}$, leaving an
exploitable un-touched fraction.}
\label{tab:design-space}
\end{center}

\par\medskip\noindent\begin{minipage}{\linewidth}\centering
\setlength{\fboxsep}{0pt}\setlength{\fboxrule}{0.4pt}
% Each panel inlined as a minipage so arXiv's source scanner sees every
% \includegraphics path literally (a \newcommand-based factory hides the
% paths behind a parameter and breaks arXiv's missing-file check).
% \dspw = the image width that leaves room for the fbox rule inside a
% 0.22\textwidth minipage (computed once with \dimexpr; \includegraphics
% keyval cannot parse \dimexpr in its options).
\newlength{\dspw}\setlength{\dspw}{\dimexpr 0.22\textwidth - 0.8pt\relax}
\begin{minipage}[t]{0.22\textwidth}\centering
  \fbox{\includegraphics[width=\dspw]{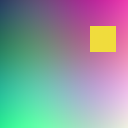}}\\{\scriptsize original}
\end{minipage}\hfill
\begin{minipage}[t]{0.22\textwidth}\centering
  \fbox{\includegraphics[width=\dspw]{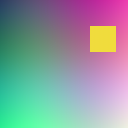}}\\{\scriptsize RGB 255}
\end{minipage}\hfill
\begin{minipage}[t]{0.22\textwidth}\centering
  \fbox{\includegraphics[width=\dspw]{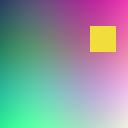}}\\{\scriptsize RGB 128 (7-bit)}
\end{minipage}\hfill
\begin{minipage}[t]{0.22\textwidth}\centering
  \fbox{\includegraphics[width=\dspw]{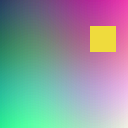}}\\{\scriptsize \textbf{RGB 64 (default)}}
\end{minipage}\\[1ex]
\begin{minipage}[t]{0.22\textwidth}\centering
  \fbox{\includegraphics[width=\dspw]{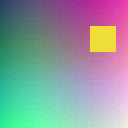}}\\{\scriptsize RGB 32 (5-bit)}
\end{minipage}\hfill
\begin{minipage}[t]{0.22\textwidth}\centering
  \fbox{\includegraphics[width=\dspw]{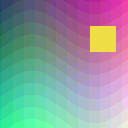}}\\{\scriptsize RGB 16 (4-bit)}
\end{minipage}\hfill
\begin{minipage}[t]{0.22\textwidth}\centering
  \fbox{\includegraphics[width=\dspw]{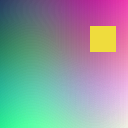}}\\{\scriptsize CMYK 64}
\end{minipage}\hfill
\begin{minipage}[t]{0.22\textwidth}\centering
  \fbox{\includegraphics[width=\dspw]{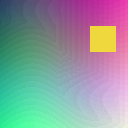}}\\{\scriptsize CMYK 32}
\end{minipage}\\[1ex]
\begin{minipage}[t]{0.22\textwidth}\centering
  \fbox{\includegraphics[width=\dspw]{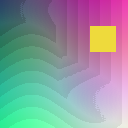}}\\{\scriptsize K-only 4-bit}
\end{minipage}\hfill
\begin{minipage}[t]{0.22\textwidth}\centering
  \fbox{\includegraphics[width=\dspw]{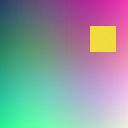}}\\{\scriptsize CMYK rand [6,7]}
\end{minipage}\hfill
\begin{minipage}[t]{0.22\textwidth}\centering
  \fbox{\includegraphics[width=\dspw]{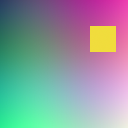}}\\{\scriptsize RGB rand 1-of-3}
\end{minipage}\hfill
\begin{minipage}[t]{0.22\textwidth}\centering
  \fbox{\includegraphics[width=\dspw]{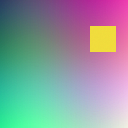}}\\{\scriptsize sprinkle $p{=}0.5$}
\end{minipage}
\captionof{figure}{Design-space probe images. The same 128$\times$128 cover
under twelve representative defense settings. The cover is a
smooth \emph{additive} RGB gradient --- R, G, and B all grow
simultaneously along different axes, so the visible result is a
continuous diagonal traversal of the gamut rather than separable
red, green, and blue bands: from dark blue-grey in the top-left
(all three channels low) through magenta-ish (top-right) and
teal-ish (bottom-left) to near-white in the bottom-right where
all three channels approach 255 --- exactly the
$\max(R,G,B)=255$ / CMYK $K=0$ region where the CMYK fixed-point
escape route shows up. The small bright high-saturation yellow
patch in the upper-right quadrant is an independent sharp-edged
high-chroma reference so quantization banding at high-contrast
boundaries is visible across panels. A least-significant-bit
secret is embedded in the blue channel. The chosen default ---
RGB six bits per channel --- is visually indistinguishable from
the original at print resolution while driving both bit-0 and
bit-1 stego receivers to chance (\cref{tab:design-space}).
Aggressive settings such as \emph{RGB 16} and \emph{K-only 4-bit}
are visibly degraded (posterization, especially around the
yellow patch's edges) and included only as endpoints; subtle
settings such as \emph{RGB 255} look identical to the eye and to
the receiver ($99.4\%$ bit-0 recovery --- the channel is
essentially intact).}
\label{fig:design-space-grid}
\end{minipage}\par\medskip

The ultrasonic encoder is a binary frequency-shift keying (BFSK)
channel: each secret bit selects one of two ultrasonic carrier
frequencies (above $20$\,kHz), and the receiver demodulates by
band-pass-filtering each frequency and reading the active one per
symbol. The sub-perceptual encoder hides each bit as a narrowband
tone $60$\,dB below the loudest audible bin.

\begin{center}
\small
\begin{tabular}{@{}lrrr@{}}
\toprule
Audio carrier & Defense & Reduction & Receiver acc.\ \\
\midrule
ultrasonic ($>20$ kHz)    & none           &   0.0\% & 100.0\% \\
ultrasonic ($>20$ kHz)    & FFT band-limit & 100.0\% &  52.7\% \\
sub-perceptual ($-60$ dB) & none           &   0.0\% &  99.4\% \\
sub-perceptual ($-60$ dB) & perceptual floor & 100.0\% & 51.1\% \\
\bottomrule
\end{tabular}
\captionof{table}{Out-of-band and sub-perceptual audio carriers
against the audio scrambler. The FFT pass removes both, with the
audible band preserved.}
\label{tab:audio}
\end{center}

\paragraph{Visualising the spectral defense.}
\cref{fig:audio-spectra} shows what the audio scrambler does to
a contaminated signal, in the frequency domain. The clean
carrier (panel~a) is a major chord at 440, 554, and 659\,Hz; the
contaminated version (panel~b) is the same chord plus two covert
components --- a binary frequency-shift keying (BFSK) ultrasonic
channel switching between $f_0=21.5$\,kHz and $f_1=22.5$\,kHz at
a symbol rate of $50$\,baud, and a sub-perceptual narrowband tone
at 5\,kHz, $60$\,dB
below the loudest audible bin. The scrambled output (panel~c) is
what the host's media-egress chokepoint would emit: the audible
chord is bit-identical to the clean version while both covert
components are zeroed. The two ultrasonic BFSK peaks --- each at
about $-15$\,dB in the contaminated spectrum --- are gone
entirely (they lie outside the $[20, 20{,}000]$\,Hz audible band
passed by the scrambler); the 5\,kHz sub-perceptual tone is gone
because it sits below the perceptual floor relative to the
chord's loudest bin and the scrambler zeroes every bin below that
threshold. Concretely, on this single $170$\,ms clip the
scrambler zeroed $1{,}372$ out-of-band bins and $5{,}792$
sub-perceptual bins.

\begingroup
\centering
\pgfplotsset{
  every axis/.style={
    width=0.92\textwidth, height=3.2cm,
    xmode=log, xmin=20, xmax=24000,
    ymin=-100, ymax=5,
    xlabel={frequency (Hz, log)},
    ylabel={magnitude (dB)},
    xlabel style={font=\footnotesize},
    ylabel style={font=\footnotesize},
    tick label style={font=\scriptsize},
    title style={font=\small, yshift=-2pt},
    xtick={20,100,1000,10000,20000},
    xticklabels={20,100,1k,10k,20k},
    grid=major, grid style={gray!20, very thin},
    enlarge x limits=false,
  },
}
\begin{tikzpicture}
\begin{axis}[title={(a)~clean: three-tone audible chord (440, 554, 659\,Hz)}]
  \addplot[blue!70!black, thin] table {figures/audio-spectrum-clean.dat};
\end{axis}
\end{tikzpicture}

\vspace{0.5em}

\begin{tikzpicture}
\begin{axis}[title={(b)~contaminated: chord + ultrasonic BFSK ($f_0{=}21.5{\rm k}$, $f_1{=}22.5{\rm k}$) + 5\,kHz sub-perceptual}]
  \addplot[red!70!black, thin] table {figures/audio-spectrum-contaminated.dat};
  % Annotate the two covert peaks.
  \draw[->, gray, thin] (axis cs:15000,-30) -- (axis cs:21500,-20)
    node[pos=0, above left, font=\footnotesize, gray] {BFSK $f_0$};
  \draw[->, gray, thin] (axis cs:18000,-45) -- (axis cs:22500,-20)
    node[pos=0, below left, font=\footnotesize, gray] {BFSK $f_1$};
  \draw[->, gray, thin] (axis cs:2200,-65) -- (axis cs:5000,-60)
    node[pos=0, above left, font=\footnotesize, gray] {sub-perceptual};
\end{axis}
\end{tikzpicture}

\vspace{0.5em}

\begin{tikzpicture}
\begin{axis}[title={(c)~scrambled by \texttt{egress-audio-scrambler}: covert content gone, audible chord intact}]
  \addplot[green!50!black, thin] table {figures/audio-spectrum-scrambled.dat};
\end{axis}
\end{tikzpicture}

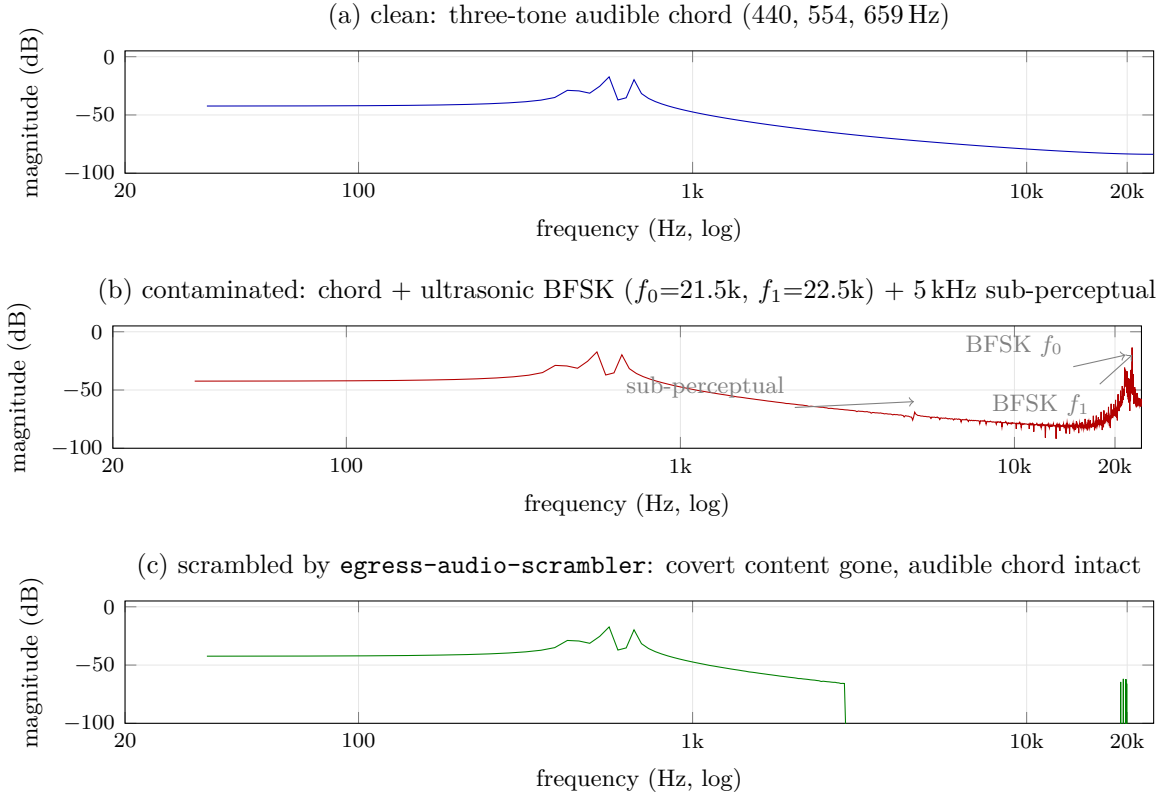
\captionof{figure}{Audio-scrambler spectral defense in action. Magnitude
spectra (dB relative to the loudest bin of the clean signal,
log-frequency axis) of (a)~a clean three-tone audible chord,
(b)~the same chord contaminated with a binary frequency-shift-
keying (BFSK) ultrasonic channel ($f_0=21.5$\,kHz / $f_1=22.5$\,kHz
at a symbol rate of $50$\,baud, peaks at $-15$\,dB each) and a 5\,kHz sub-perceptual
narrowband stego tone $60$\,dB down, and (c)~the same contaminated
signal
after the \texttt{egress-audio-scrambler} fast-Fourier-transform
pass. The audible chord components at 440, 554, and 659\,Hz pass
through unchanged in both contaminated and scrambled spectra;
the two covert components present in panel (b) are absent from
panel (c) because the scrambler zeroes every bin outside
$[20, 20{,}000]$\,Hz and every bin below a perceptual floor
relative to the loudest bin. What a human hears is preserved;
what only a covert receiver could demodulate is gone.}
\label{fig:audio-spectra}
\par\endgroup\vspace{0.6em}

\subsection{The audible-band sonification case}

The audible-band sonification carrier of \cref{tab:carriers} is
the one this paper does not attempt to defeat with a content-based
transform. Its treatment under the design is the legitimacy
attestation of \cref{sec:legitimacy}: unsigned audible audio is
scrambled (the FFT pass strips its out-of-band and sub-perceptual
content, which already removes the precise spectral structure
many sonification encodings rely on; we observe full reduction on
the FSK-style audible-tone encoders we tried but make no general
claim, because audible-band sonification is an open arms race);
signed legitimate audio of an authorized class is exempt.
\cref{tab:legitimacy} reports the legitimacy-gate behavior on a
small unit-level ensemble.

\begin{center}
\small
\begin{tabular}{@{}llcc@{}}
\toprule
Token state & Hash match & Class authorized & Outcome \\
\midrule
absent / unsigned        & --     & --     & scramble \\
present, signature valid & yes    & yes    & \textbf{exempt} \\
present, signature valid & yes    & no     & scramble \\
present, signature valid & no     & yes    & scramble \\
present, untrusted key   & yes    & yes    & scramble \\
present, tampered sig.\  & yes    & yes    & scramble \\
\bottomrule
\end{tabular}
\captionof{table}{Boot-time legitimacy gate, per the verifier in
\cref{sec:legitimacy}. Only the second row exempts the payload;
every other condition --- including a valid signature for an
unauthorized class, or for a different payload --- routes through
the scrambler. The decision is fail-secure when the host has not
installed a verifier (every row collapses to ``scramble'').}
\label{tab:legitimacy}
\end{center}

%% =====================================================================
\section{Limitations and future work}
\label{sec:limits}
%% =====================================================================

\paragraph{Setting the measurement standard.}
This paper proposes a standard for how covert-channel egress is
to be measured: pick a working encoder set, run each encoder
against the defense, decode whatever survives, and report the
Miller--Madow-corrected mutual information between embedded and
recovered bits. We ship fifteen encoders as a realistic starting
suite (\cref{tab:carriers}): fourteen reach zero residual
capacity and the mean-luminance encoder reaches a stated bound
(\cref{tab:cross-image}). New encoders extend the suite directly
--- each stage is a separate extension, each encoder a benchmark
row, and \cref{sec:capacity}'s measurement applies unchanged.
Which encoders to add is a concrete question: a production
deployment runs in a known domain (finance, healthcare, legal,
code review, content moderation), and the domain fixes the
vocabulary, the payload shape, and the high-entropy fields an
attacker can hide bits in. Attackers specialise to match:
credential thieves work base64 and least-significant-bit (LSB)
carriers, moderation evaders work homoglyphs and zero-width
characters, audio exfiltrators work ultrasound. The threat
surface that matters is the intersection of the deployment's
substrate and the encoders the relevant adversaries actually use;
the benchmark should be extended to cover \emph{that}. The floor
of \cref{sec:threat} remains: one bit always leaks through the
agent's choice to send or not send.

\paragraph{Text chokepoint sees text.} The mediated hook exposes
the outbound \emph{text} of a message. Structured interactive
blocks and channel-envelope metadata are not reachable at that
point and need their own chokepoints. The media scramblers of
\cref{sec:media} address the image and audio attachments, but
they presuppose a media-egress chokepoint that the host must add.

\paragraph{Bit estimates are upper bounds.} The entropy scanner's
covert-bit estimates are deliberate over-counts. Over-counting
covert capacity is the safe direction for a security control, but
it means the ledger can be conservative and the scanner can
charge benign high-entropy content (a legitimate base64
attachment, a random identifier). This is the false-positive
surface the staged posture exists to manage, and it is why the
entropy and behavioral stages audit before they enforce.

\paragraph{Paced release has a latency cost.} Constant-rate
release delays messages. For an interactive agent this is a real
user-visible cost, and it is why the strongest timing defense is
configuration, not default.

\paragraph{The LLM scrambler trusts a model.} The model-backed
rephraser sends outbound text to a language model, which adds
three costs the deterministic stages do not carry: a latency and
monetary cost, non-determinism, and a prompt-injection surface
--- the text being rephrased may itself contain instructions. The
fixed prompt mitigates the last by fencing the input strictly as
data and forbidding instruction-following, and a model failure
fails toward pass-through rather than blocking egress; but a
fixed prompt is a mitigation, not a proof, and a sufficiently
capable injection could still subvert the rephrasing. This is why
the LLM scrambler is configuration-gated and audit-default, and
why the deterministic scrambler --- which has none of these costs
--- remains the always-on linguistic defense.

\paragraph{The legitimacy attestation trusts launcher principals,
not producer code.} A signed payload is exempt from media
scrambling. The chain of \cref{sec:legitimacy-keychain} bounds
this trust in three concrete ways: keys are accreditor-minted
(not producer-generated), they are per-process ephemeral (the
window of a leaked key is one process lifetime), and they live
behind a sandbox boundary so they do not appear as raw material
in the producer's address space. None of this prevents a producer
whose CODE is subverted while it still holds its handle from
asking the sandbox to sign whatever the attacker wants while the
process runs. That residual --- a compromised but unrestarted
producer signs freely until restart --- is the one we accept; the
attestation reduces the exfiltration window to the producer's
process lifetime and pins liability to the launcher who
authorized the job, but it does not solve the
trusting-code-that-holds-a-key problem in full generality. We
believe this is the principled middle between an unworkable ``no
media'' policy and a brittle content classifier.

\paragraph{Mean luminance is bounded, not destroyed.} The
per-image mean-luminance channel cannot be perfectly destroyed at
the application layer without making the image unusable
(clamping every image to a single fixed mean defeats the purpose
of emitting it). The defense bounds it: bucketing to $B$ levels
fixes the per-image capacity at $\log_2 B$ bits. With $B=8$ that
is $3$ bits per image; with $B=2$ it is $1$ bit. The aggregate
across a sequence is then bounded by the capacity ledger's
per-sink rate. We report this as a bound, not a $100\%$
reduction.

\paragraph{Audible-band sonification is an arms race.} We claim
the legitimacy attestation closes the gap for the deployment
posture in which legitimate audio is signed and everything else is
suspect. We do not claim that any unsigned audible audio survives
the audio scrambler with its sonified payload intact; on the
FSK / amplitude-modulated encodings we tried, the FFT pass's
perceptual floor and band-limit already collapse the channel.
What we cannot rule out is a sufficiently robust sonification
that survives the spectral pass. The defense's posture for that
case is the same: unsigned, therefore suspect; the scrambler runs;
a deployment may further configure the media-egress chokepoint to
drop unsigned audio rather than scramble it.

\paragraph{The attacker is assumed outside the TCB.} If the
attacker compromises the runtime itself, the reference monitor ---
like any reference monitor --- is moot. The design defends against
a compromised \emph{agent}, not a compromised \emph{host}.

%% =====================================================================
\section{Related work and positioning}
\label{sec:related}
%% =====================================================================

This section places the contribution against six strands of prior
work and then states, concretely, what it does that none of them
do (\cref{sec:beyond}).

\paragraph{Covert channels and quantitative information flow.}
The problem is Lampson's \cite{lampson1973confinement}. Covert-
channel analysis matured in the Trusted Computer System
Evaluation Criteria (TCSEC) era \cite{tcsec1985}:
identification methodologies such as the shared-resource matrix
\cite{kemmerer1983srm}, the analysis of timing channels
specifically \cite{wray1991covert}, and the formal treatment of
\emph{capacity} \cite{shannon1948} as \emph{the} security-
relevant quantity \cite{millen1987covert}. A modern descendant,
quantitative information flow (QIF), measures the leakage of a
program as an information-theoretic quantity
\cite{smith2009qif, kopf2007adaptive}. We inherit from both
lineages the conviction that a covert-channel claim is a
\emph{capacity} claim and must be quantified. The difference is
the object of study. Classical analysis and QIF examine a system
or program one can read --- a shared hardware resource, a piece
of source code --- and reason about it statically. An LLM agent
is not such an object: its policy is an opaque, non-deterministic
model, and its egress is open-ended natural language and binary
media. We therefore do not \emph{analyze} a channel; we
\emph{deploy} a runtime monitor on the agent's egress across
modalities and \emph{measure} the monitor's residual capacity
with an adversarial encoder ensemble.

\paragraph{Information-flow control.} Lattice models
\cite{denning1976lattice, bell-lapadula}, program-certification
techniques for secure flow \cite{denning1977certification}, and
the systems that enforced them --- SELinux
\cite{loscocco2001selinux}, HiStar \cite{zeldovich2006histar} ---
propagate labels along data flows. Label propagation presupposes
a tractable flow graph. The agent's reasoning is not one: data
enters the model's context and re-emerges transformed in ways no
taint engine can follow. Our taint tracker is therefore
deliberately not whole-program information-flow control (IFC); it tags sensitive values at
ingress and detects them at the egress chokepoint, and ---
crucially --- it is backed by the capacity ledger so that
\emph{unknown} encodings of sensitive data, which label tracking
would never catch once they have passed through the model, are
still bounded.

\paragraph{Linguistic and multi-modal steganography.} A mature
line of work shows how to hide payloads inside natural-language
text, increasingly inside LLM-generated text: neural linguistic
steganography \cite{ziegler2019neural} and, most pointedly,
constructions that are \emph{provably} secure --- statistically
undetectable --- against any steganalysis
\cite{dewitt2023steganography}. The bitmap and audio literatures
are equally mature: LSB steganalysis on images
\cite{fridrich2001steganalysis} is a foundational defensive
result, and ultrasonic and sub-perceptual covert channels in
audio are demonstrated attacks
\cite{deshotels2014inaudible}. Sonification itself is a
legitimate research area \cite{hermann2011sonification} whose
techniques double as covert carriers. This is the attacker's
literature, and it carries a hard lesson for the defender:
detection alone is a losing strategy, because the strongest stego
is information-theoretically undetectable by construction. That
lesson is the reason our design does not rest on detecting a
particular hiding scheme. The canonicalizer and noise injector
\emph{destroy the carrier's capacity} regardless of scheme; the
entropy scanner's role is to \emph{budget}, not to classify; the
image and audio scramblers destroy the carrier's binary capacity;
and for audible-band sonification --- where neither destruction
nor detection works without collateral damage --- the legitimacy
attestation makes signature, not classification, the
discriminator. We are, to our knowledge, the first to draw the
defensive consequence of provably-secure steganography and
audible-band sonification for an agent runtime: if you cannot
detect it and you cannot safely destroy it, you must make
legitimacy explicit and cryptographic, and treat the absence of
a signature as the suspect default.

\paragraph{Text-carrier attacks.} Trojan Source
\cite{boucher2023trojansource} demonstrated bidirectional-control
and homoglyph attacks on source code; Unicode Tag-character
``ASCII smuggling'' \cite{ascii-smuggling} demonstrated invisible
instruction and data carriers against LLM applications
specifically. Each came with a point mitigation --- scan for bidi
controls; strip tag characters. The prevailing practice is
exactly this: one ad hoc fix per carrier as it is discovered.
The canonicalizer subsumes the known carriers of this class, but
the structural contribution is that it is one stage of a measured
pipeline rather than a standalone scanner, so a newly disclosed
carrier is added to a benchmark with a capacity number, not
shipped as another disconnected patch.

\paragraph{Timing channels and traffic shaping.} Covert timing
channels and their detection are long studied
\cite{wray1991covert, cabuk2004timing}. The traffic-analysis
literature contributes a sharper, and sobering, result: Dyer et
al.\ \cite{dyer2012peekaboo} showed that \emph{efficient} timing
countermeasures --- including jitter and adaptive padding ---
leak, and that only genuinely constant-rate transmission defeats
a determined analyst. This is precisely why our behavioral stage
does not stop at quantization-plus-jitter (which \cref{sec:eval}
indeed measures as leaky) but provides constant-rate paced
release, and why the paced-release result is the one we report as
closing the channel.

\paragraph{LLM guardrails, DLP, and agent exfiltration.}
Model-boundary filters \cite{inan2023llamaguard} classify inputs
and outputs for policy-violating content; indirect prompt
injection \cite{greshake2023indirect} is the demonstrated
mechanism by which a third party hijacks an agent's egress
intent --- the attacker of our threat model; data-loss-prevention
systems \cite{shabtai2012dlp} match egress against signatures of
known-sensitive data; and the OWASP LLM Top 10
\cite{owasp-llm-top-10} and MITRE ATLAS \cite{mitre-atlas}
catalogue agent exfiltration as a first-class risk. Every one of
these asks a \emph{content} question --- is this output
disallowed, does it contain a known secret. None asks the
\emph{channel} question: is this allowed-looking, secret-free
output nonetheless carrying bits. A guardrail and a DLP scanner
both pass a payload that encodes a key in its whitespace; the
egress monitor does not. Neither asks the \emph{legitimacy}
question for media: is this audio a legitimate TTS rendering or
is it the same waveform produced by sonifying a bitstream? The
monitor is complementary to all of them and sits, deliberately,
in the runtime rather than in the model, independent of the
model's own classification of its text
\cite{metere2026enclawed}.

\subsection{Beyond the state of the art}
\label{sec:beyond}

Against that landscape, this paper makes six advances that, to
our knowledge, no prior system or study combines --- or in
several cases attempts at all.

\begin{enumerate}[leftmargin=1.6em]
\item \textbf{A covert-channel reference monitor for agent
linguistic egress.} Covert-channel discipline has, for fifty
years, been applied to hardware and operating-system resources;
LLM defenses have been content classifiers. We are the first,
to our knowledge, to treat an agent's natural-language and
structured egress as a covert-channel medium and to place a
\emph{capacity-reducing} reference monitor --- not a classifier
--- on it.

\item \textbf{Multi-modal coverage with the same metric.} The
LLM-security literature is wholly text-focused; the bitmap and
audio covert-channel literatures are entirely siloed from agent
egress. We treat text, image, and audio as one egress surface
with one mutual-information benchmark, and measure all of them
the same way.

\item \textbf{Boot-time cryptographic legitimacy attestation as
the false-positive escape valve for media.} Application-layer
content classification cannot robustly distinguish real media
from data sonified or rasterized into the same byte range
without unacceptable false-positive rates. We are, to our
knowledge, the first to propose a boot-time-only auditor model
in which the deployment declares its authorized data kinds and
signing keys, legitimate producers sign their outputs, and the
media scramblers exempt validly-signed payloads of an authorized
class --- so the discriminator is cryptographic, the runtime
check is a signature verification, and the false-positive
question reduces to key trust rather than to spectral or
statistical content classification.

\item \textbf{Capacity, measured, as the evaluation paradigm.}
The norm in LLM security is to report attack-success-rate or
detection F1 on a prompt corpus. Those numbers cannot answer
``how many bits leak.'' We report the information-theoretically
correct quantity: Miller-Madow-corrected mutual information
between embedded and recovered bits, measured by an adversarial
ensemble of \emph{working} encoders and receivers across three
modalities. This is a shift in what an egress-defense claim is
allowed to mean, and it makes the claim falsifiable: a reader
can add an encoder and rerun.

\item \textbf{A unified, capacity-budgeted pipeline.} Where the
field ships one disconnected mitigation per carrier, we give a
single mediated chokepoint running an ordered detect / disrupt /
constrain pipeline, tied together by a per-sink capacity ledger
that converts the scanner's measurement into the rate-limiter's
enforcement. Adding a defense is adding a stage; adding an
attack is adding a benchmark row.

\item \textbf{Side channels in scope, and closed.} Timing, size,
rate, and ordering are essentially absent from the LLM-security
literature, which is wholly content-focused. We bring them into
a single egress defense and, taking the traffic-analysis
negative result \cite{dyer2012peekaboo} seriously, close the
timing channel with a measured constant-rate paced-release
result rather than the jitter that prior practice would suggest
and that we show to be insufficient.
\end{enumerate}

\noindent The question of covert egress in an agent runtime is
now asked at the right layer, across three modalities, answered
with the right metric, and shipped with a benchmark that lets the
answer be checked and extended.

%% =====================================================================
\section{Conclusion}
\label{sec:conclusion}
%% =====================================================================

Destination control and content control leave a gap, and in an
LLM agent runtime that gap is wide and now multi-modal: an agent
emits free-form text constantly, and --- increasingly ---
images and audio. Each is a rich covert carrier; the third, audio,
admits a carrier (audible-band sonification) that no application-
layer content classifier can robustly tell from real audio. We
have argued that the gap should be closed by a single mediated
egress chokepoint per modality, running an ordered, capacity-
budgeted pipeline of detect / disrupt / constrain techniques for
text and a pair of media scramblers gated by a boot-time
cryptographic legitimacy attestation for binary, with a staged
posture so the text control can be deployed without breaking
legitimate traffic on day one and a signed-or-suspect default so
the media control cannot silently degrade real graphics or audio.
We have also argued --- and this is the part we would most like
to see adopted --- that a covert-channel control is only a
security claim if it ships with a capacity measurement, that the
correct measurement is mutual information between embedded and
recovered bits, and that the measurement belongs in an
adversarial benchmark of working encoders across every modality
the defense touches. The reference implementation drives measured
residual capacity to zero across the tested text, image, and
audio ensemble under full enforcement for every destroyable
channel, and to a stated bound for the one cross-image channel
(mean luminance) that it can only bound. The benchmark, not the
number, is what we hope generalizes. The legitimacy attestation
moves the trust problem to producer-side key hygiene rather than
dissolving it; that, and the open multi-modal benchmark, are what
we offer as durable contributions.

\paragraph{Availability.} Every mechanism described in this paper
--- the egress reference monitor, the ten text-domain stages with
their capacity ledger and staged-enforcement posture, the
Fourier-domain audio scrambler, the RGB-basis image-bit-depth and
mean-luminance scrambler, the content-addressed media-sequence
canonicalizer, the boot-time cryptographic media legitimacy
attestation with its accreditor-brokered launcher $\to$ identity
$\to$ signing-key chain and sandbox-bound handles, and the
adversarial mutual-information benchmark harness --- is
\emph{already integrated and production-ready} in Enclawed
\cite{metere2026enclawed}, and is the reference implementation
against which every figure in this paper is measured.
Deployments enable the suite by publishing the auditor's
boot-time media-legitimacy trust set; no further integration is
required for the text chokepoint, and the media-egress
chokepoint is the single host seam the deployment must wire to
its outbound media surface.

\small
\bibliographystyle{plainnat}
\bibliography{refs}

\end{document}